\documentclass[a4paper]{article}
\usepackage{amssymb}
\newcommand{\bea}{\begin{eqnarray}}
\newcommand{\eea}{\end{eqnarray}}
\newcommand{\ba}{\begin{array}}
\newcommand{\ea}{\end{array}}
\newcommand{\bc}{\begin{center}}
\newcommand{\ec}{\end{center}}
\newcommand{\be}{\begin{equation}}
\newcommand{\ee}{\end{equation}}

\newcommand{\dsf}{\displaystyle\frac}
\def\s{\sigma}

\def\Q{\mathbb{Q}}
\def\Z{\mathbb{Z}}

\def\O{\Omega}

\def\m{\mu}

\def\g{\gamma}

\begin{document}

\begin{center}
{\bf ON PHASE TRANSITIONS FOR $P$-ADIC POTTS MODEL WITH COMPETING
INTERACTIONS ON A  CAYLEY
TREE}\\[1.4cm]
FARRUKH  MUKHAMEDOV\\[1mm]
\small {\it
Department of Mechanics and Mathematics, \\
National University of Uzbekistan,\\
Vuzgorodok, 700174,  Tashkent, Uzbekistan\\
E-Mail: far75m@yandex.ru}\\[4mm]
UTKIR A. ROZIKOV\\[1mm]
\small {\it Institute of Mathematics, 29,  F.Hodjaev str.,
Tashkent, 700143,
Uzbekistan\\
E-mail: root@im.tashkent.su}\\[1.5cm]
\end{center}
\begin{abstract}
We consider a nearest-neighbor $p$-adic Potts (with $q\geq 2$ spin
values and coupling constant $J\in \Q_p$) model on the Cayley tree
of order $k\geq 1$. It is proved  that a phase transition occurs
at $k=2$, $q\in p\mathbb{N}$ and $p\geq 3$ (resp. $q\in
2^2\mathbb{N}$, $p=2$). It is established that for $p$-adic Potts
model at $k\geq 3$ a phase transition may occur only at $q\in
p\mathbb{N}$ if $p\geq 3$ and $q\in 2^2\mathbb{N}$ if $p=2$.

{\it Keywords:} $p$-adic field, Potts model, Cayley tree, Gibbs
measure, phase transition

{\it AMS Subject Classification:} 46S10, 82B26, 12J12.
\end{abstract}


\section{Introduction}

The $p$-adic numbers were first introduced by the German
mathematician K.Hensel. For about a century after the discovery of
$p$-adic numbers, they were mainly considered objects of pure
mathematics. However, numerous applications of these numbers to
theoretical physics have been proposed papers [1-5] to quantum
mechanics [6], to $p$-adic - valued physical observables [6] and
many others [7,8]. A number of $p$-adic models in physics cannot
be described using ordinary probability theory based on the
Kolmogorov axioms [9]. New probability models - $p$-adic
probability models
 were investigated in [8],[10].

In [11,12] a theory of stochastic processes with values  in
$p$-adic and more general non-Archimedean fields was developed ,
having probability distributions with non-Archimedean values.

One of the basic branches of mathematics lying  at the base of the
theory of statistical mechanics is the theory of probability and
stochastic processes. Since  the theories of probability and
stochastic processes in a non-Archimedean setting have been
introduced, it is natural to study  problems of statistical
mechanics in the context of the $p$-adic theory of probability.

Many physical models are considered on $\Z^d$ and the Cayley tree.
The difference between these two graphs is that the Cayley tree
has the property that the number of neighbors visited in $n$ steps
grows exponentially with $n$. This is a faster rate of growth than
$n^d$, no matter how large $d$ is, so this tree is infinite
dimensional.

In this paper we develop the $p$-adic probability theory approach
to study some statistical mechanics models on a Cayley tree over
the field of $p$-adic numbers. In [13] the Potts model with $q$
spin variables on the set of integers $\mathbb{Z}$ in the field of
$\Q_p$ was studied. It is known [14] that for the Potts model and
even for arbitrary models on $\mathbb{Z}$ regardless of the
interaction radius of the particles (over  $\mathbb{R}$) there are
no phase transitions, here the phase transition means that for
given Hamiltomian there are at least two  Gibbs measures. In the
case considered in [13] this pattern is destroyed, namely,
 there are some
values $q=q(p)$ for which phase transition occurs.

In the present paper we consider $p$-adic Potts models ( with
coupling constant $J$ and $q$ spin variables) on the Cayley tree
of order $k$, $k\geq 1$. The aim of this paper is to investigate
Gibbs measures for $p$-adic Potts model and  phase transition
problem for this model. The organization of this paper as follows.

Section 2 is a mathematically preliminary. In section 3 we give a
construction of Gibbs measure for the $p$-adic Potts model on the
Cayley tree. In section 4 we prove the existence of a phase
transition for $p$-adic Potts model on Cayley tree of order two.
In final section 5 we exhibit some conditions implying $q$  on
uniqueness of the Gibbs measure.

\section{Definitions and preliminary results}

\subsection{$p$-adic numbers and measures}

Let $\Q$ be the field of rational numbers. Every rational number
$x\neq 0$ can be represented in the form $x=p^r\dsf{n}{m}$, where
$r,n\in\mathbb{Z}$, $m$ is a positive integer, $(p,n)=1$,
$(p,m)=1$ and $p$ is a fixed prime number. The $p$-adic norm of
$x$ is given by
$$
|x|_p=\left\{ \ba{ll}
p^{-r} & \ \textrm{ for $x\neq 0$}\\
0 &\ \textrm{ for $x=0$}.\\
\ea \right.
$$
This norm satisfies so called the strong triangle inequality
$$
|x+y|_p\leq\max\{|x|_p,|y|_p\},
$$
this is a non-Archimedean norm.

The completion of $\Q$ with  respect to $p$-adic norm defines the
$p$-adic field which is denoted by $\Q_p$. Any $p$-adic number
$x\neq 0$ can be uniquely represented in the form  $$
x=p^{\g(x)}(x_0+x_1p+x_2p^2+...) , \eqno(2.1) $$ where
$\g=\g(x)\in\Z$ and $x_j$ are integers, $0\leq x_j\leq p-1$,
$x_0>0$, $j=0,1,2,...$ (see more detail [7,15]). In this case
$|x|_p=p^{-\g(x)}$.

We recall that an integer $a\in \Z$ is called {\it a quadratic
residue modulo $p$} if the equation $x^2\equiv a(\textrm{mod
$p$})$ has a solution $x\in \Z$.

{\bf Lemma 2.1.}[7] {\it In order that the equation
$$
x^2=a, \ \ 0\neq a=p^{\g(a)}(a_0+a_1p+...), \ \ 0\leq a_j\leq p-1,
\ a_0>0
$$
has a solution $x\in \Q_p$, it is necessary and sufficient that
the following conditions are fulfilled:

i) $\g(a)$ is even;

ii) $a_0$ is a quadratic residue modulo $p$ if $p\neq 2$, $a_1=a_2=0$ if $p=2$.}\\

Let $B(a,r)=\{x\in \Q_p : |x-a|_p< r\}$, where $a\in \Q_p$, $r>0$.
By $\log_p$ and $\exp_p$ we mean $p$-adic logarithm and
exponential which are defined as series with the usual way (see,
for more details [7]. The domain of converge for them are $B(1,1)$
and $B(0,p^{-1/(p-1)})$ respectively.)

{\bf Lemma 2.2.}[7,15,16] {\it Let $x\in B(0,p^{-1/(p-1)})$ then
we have $$ |\exp_p(x)|_p=1,\ \ \ |\exp_p(x)-1|_p=|x|_p<1, \ \
|\log_p(1+x)|_p=|x|_p<p^{-1/(p-1)} $$ and $$ \log_p(\exp_p(x))=x,
\ \ \exp_p(\log_p(1+x))=1+x. $$ }

Let $(X,{\cal B})$ be a  space, where ${\cal B}$ is an algebra of
subsets $X$. A function $\m:{\cal B}\to \Q_p$ is said to be a {\it
$p$-adic measure} if for any $A_1,...,A_n\subset{\cal B}$ such
that $A_i\cap A_j=\emptyset$ ($i\neq j$)

$$
\mu(\bigcup_{j=1}^{n} A_j)=\sum_{j=1}^{n}\mu(A_j).
$$

A $p$-adic measure is called {\it a probability measure} if
$\mu(X)=1$. A $p$-adic probability measure $\m$ is called {\it
bounded} if $\sup\{|\m(A)|_p : A\in {\cal B}\}<\infty $.

For more detail information about $p$-adic measures we refer to
[8],[10].

\subsection{The Cayley tree}

The Cayley tree  $\Gamma^k$ of order $ k\geq 1 $ is an infinite
tree, i.e., a graph without cycles, such that each vertex of which
lies on $ k+1 $ edges . Let $\Gamma^k=(V, \Lambda),$  where $V$ is
the set of vertices of $ \Gamma^k$, $\Lambda$ is the set of edges
of $ \Gamma^k$. The vertices $x$ and $y$ are called {\it nearest
neighbor}, which is denoted by $l=<x,y>$ if there exists an edge
connecting them. A collection of the pairs
$<x,x_1>,...,<x_{d-1},y>$ is called {\it a path} from  $x$ to $y$.
The distance $d(x,y), x,y\in V$  is the length of the shortest
path from $x$ to $y$ in V.

We set
$$ W_n=\{x\in V| d(x,x^0)=n\}, $$
$$ V_n=\cup_{m=1}^n W_m=\{x\in V| d(x,x^0)\leq n\}, $$
$$ L_n=\{l=<x,y>\in L | x,y\in V_n\}, $$
for a fixed point $ x^0 \in V $.

We write  $x<y$  if the path from $x^0$ to $y$ goes through $x$.
Call vertex $y$ a {\it direct successor} of $x$ if $y>x$ and $x,y$
are nearest neighbors. Denote by $S(x)$ the set of direct
successors, i.e.
$$
S(x)=\{y\in W_{n+1} :  d(x,y)=1 \} \ \ x\in W_n,
$$
Observe that any vertex $x\neq x^0$ has $k$ direct successors and
$x^0$ has $k+1$.

\subsection{The $p$-adic Potts model}

Let $\Q_p$ be the field of $p$-adic numbers. By $\Q_p^{q-1}$ we
denote $\underbrace{\Q_p\times...\times\Q_p}_{q-1}$. The norm
$\|x\|_p$ of an element $x\in \Q_p^{q-1}$ is defined by
$\|x\|_p=\max\limits_{1\leq i\leq q-1}\{|x_i|_p\}$, here
$x=(x_1,...,x_{q-1})$. By $xy$ we mean the bilinear form on
$\Q_p^{q-1}$ defined by
$$
xy=\sum_{i=1}^{q-1}x_iy_i, \ \ x=(x_1,\cdots,x_{q-1}),
y=(y_1,\cdots,y_{q-1}).
$$

Let $\Psi=\{\s_1,\s_2,...,\s_q\}$, where $\s_1,\s_2,...,\s_q$ are
elements of $\Q_p^{q-1}$ such that  $\|\s_i\|_p=1$, $i=1,2,...,q$
and
$$
\s_i\s_j= \left\{ \ba{ll}
1, \ \  \textrm{for $i=j$},\\
0, \ \ \textrm{for $i\neq j$}\\
\ea \right. (i,j=1,2,...,q-1), \ \s_q=\sum_{i=1}^{q-1}\s_i.
\eqno(2.2)
$$

Let $h\in \Q_p^{q-1}$, then we have $h=\sum_{i=1}^{q-1}h_i\s_i$
and
$$
h\s_i= \left\{ \ba{ll}
h_i, \ \ \textrm{for $i=1,2,...,q-1$},\\
\sum_{i=1}^{q-1}h_i, \ \ \textrm{for $i=q$}\\
\ea \right. \eqno(2.3)
$$

We consider the $p$-adic Potts model where spin takes values in
the set $\Psi$. Write $\O_n=\Psi^{V_n}$, this is the configuration
space on $V_n$. The Hamiltonian $H_n:\O_n\to\Q_p$ of the $p$-adic
Potts model has the form $$ H_n(\s)=-J\sum_{<x,y>\in
L_n}\delta_{\s(x),\s(y)},  \ \ n\in\mathbb{N}, \eqno(2.4) $$ here
$\s=\{\s(x) : x\in V_n\}\in\O_n$, $|J|_p<p^{-1/(p-1)},$ $J\neq 0$
and as before $p$ is a fixed prime number.  Here $\delta$ is the
Kronecker symbol.

\section{Construction  of Gibbs measures}

In this section we give a  construction of a special class of
Gibbs measures for $p$-adic  Potts models on the Cayley tree.

To define Gibbs measure we need in the following

{\bf Lemma 3.1.} {\it Let $h_x, x\in V$ be a $\Q_p^{q-1}$-valued
function  such that $\|h_x\|_p\in B(0,p^{-1/(p-1)})$ for all $x\in
V$ and $J\in B(0,p^{-1/(p-1)})$. Then $$ H_n(\s)+\sum_{x\in
W_n}h_x\s(x)\in B(0,p^{-1/(p-1)}) $$ for any $n\in \mathbb{N}$.}

The proof easily follows from the strong triangle inequality for
the norm $|\cdot|_p$.

Let $h:x\in V\to h_x\in\Q_p^{q-1}$ be a function of $x\in V$ such
that $\|h_x\|_p<p^{-1/(p-1)}$ for all $x\in V$. Given $n=1,2,...$
consider a $p$-adic probability measure $\m^{(n)}$ on $\Psi^{V_n}$
defined by
$$
\mu^{(n)}(\s_n)=Z^{-1}_{n}\exp_p\{-H_n(\s_n)+\sum_{x\in
W_n}h_x\s(x)\}, \eqno(3.1)
$$
Here, as before, $\s_n:x\in V_n\to\s_n(x)$ and $Z_n$ is the
corresponding partition function:
$$
Z_n=\sum_{\tilde\s_n\in\Omega_{V_n}}\exp_p\{-H(\tilde\s_n)+\sum_{x\in
W_n}h_x\tilde\s(x)\}.
$$

Note that according to Lemma 3.1 the measures $\m^{(n)}$ exist.

The compatibility condition for $\m^{(n)}(\s_n), n\geq 1$ are
given by the equality
$$
\sum_{\s^{(n)}}\m^{(n)}(\s_{n-1},\s^{(n)})=\m^{(n-1)}(\s_{n-1}),
\eqno(3.2)
$$
where $\s^{(n)}=\{\s(x), x\in W_n\}$.

We note that an analog of the Kolmogorov extension theorem for
distributions can be proved for $p$-adic distributions given by
(3.1) (see [12]). According to this  theorem there exists a unique
$p$-adic measure $\m_h$ on $\O=\Psi^V$ such that for every
$n=1,2,...$ and $\s_n\in\Psi^{V_n}$
$$
\m\bigg(\{\s|_{V_n}=\s_n\}\bigg)=\m^{(n)}(\s_n).
$$
$\m_h$ will be called {\it $p$-adic Gibbs measure} for this Potts
model. It is clear that the measure $\m_h$ depends on function
$h_x.$ If the Gibbs measure for a given Hamiltonian is non unique
then we say that for this model there is {\it a phase transition}.

The following statement describes conditions on $h_x$ guaranteeing
the compatibility condition of measures $\m^{(n)}(\s_n)$.

{\bf Theorem 3.2.} {\it The measures $\m^{(n)}(\s_n), \ n=1,2,...$
satisfy the compatibility condition  (3.2) if and only if for any
$x\in V$ the following equation holds: $$ h_x=\sum_{y\in
S(x)}F(h_y;\theta,q) \eqno(3.3) $$ here and below
$\theta=\exp_p(J)$ and the function $F:\Q_p^{q-1}\to\Q_p^{q-1}$
function is defined by
$F(h;\theta,q)=(F_1(h;\theta,q),...,F_q(h;\theta,q))$ with $$
F_i(h;\theta,q)=\sum_{j=1,j\neq i}^{q-1}G_j(h';\theta,q) \ \
i=1,...,q-1, $$ where $h=(h_1,...,h_{q-1}),$
$h'=(h_1',...,h_{q-1}')$ $h_i'=\sum\limits_{j=1,j\neq
i}^{q-1}h_j$, $i=1,...,q-1$, $$ G_i(h_1,...,h_{q-1};\theta,q)=
\log_p\bigg[\frac{(\theta-1)\exp_p(h_i)+\sum_{j=1}^{q-1}\exp_p(h_j)+1}
{\sum_{j=1}^{q-1}\exp_p(h_j)+\theta}\bigg], $$ $i=1,...,q-1$.}

{\bf Proof.} {\it Necessity.} According to the compatibility
condition (3.2) we have $$
Z^{-1}_{n}\sum_{\s^{(n)}}\exp_p\bigg[J\sum_{<x,y>\in
L_n}\delta_{\s(x),\s(y)}+\sum_{x\in W_n}h_x\s(x)\bigg]=$$ $$
Z^{-1}_{n-1}\exp_p\bigg[J\sum_{<x,y>\in
L_{n-1}}\delta_{\s(x),\s(y)}+\sum_{x\in
W_{n-1}}h_x\s(x)\bigg].\eqno(3.4)$$ It yields $$
\frac{Z_{n-1}}{Z_n}\sum_{\s{(n)}}\exp_p\bigg[J\sum_{x\in
W_{n-1}}\sum_{y\in S(x)}\delta_{\s(x),\s(y)}+\sum_{x\in
W_{n-1}}\sum_{y\in S(x)}h_y\s(y)\bigg]=$$ $$\prod_{x\in
W_{n-1}}\exp_p\bigg(h_x\s(x)\bigg). \eqno(3.5) $$ From this
equality we find $$ \frac{Z_{n-1}}{Z_n}\prod_{x\in
W_{n-1}}\prod_{y\in S(x)}\sum_{\s(y)\in\Psi}
\exp_p\bigg(J\delta_{\s(x),\s(y)}+h_y\s(y)\bigg)=\prod_{x\in
W_{n-1}}\exp_p\bigg(h_x\s(x)\bigg).\eqno(3.6) $$ Now fix $x\in
W_{n-1}$.  Dividing the equalities (3.6) with $\s(x)=\s_i$ and
with $\s(x)=\s_q$ we obtain $$ \prod_{y\in
S(x)}\frac{\sum_{\s(y)\in\Psi}
\exp_p\bigg(J\delta_{\s_i,\s(y)}+h_y\s(y)\bigg)}{\sum_{\s(y)\in\Psi}
\exp_p\bigg(J\delta_{\s_q,\s(y)}+h_x\s(y)\bigg)}=\exp_p\bigg(h_x(\s_i-\s_q)\bigg).
\eqno(3.7) $$

 Using (2.3) the last equality can be rewritten as

$$
\prod_{y\in S(x)}\frac{\sum_{m=1}^{q-1}\exp_p\bigg(\sum_{j=1,j\neq
m}^{q-1}h_{x}^{(j)}\bigg) +(\theta-1)\exp_p\bigg(\sum_{j=1,j\neq
i}^{q-1}h_{x}^{(j)}\bigg)+
1}{\sum_{m=1}^{q-1}\exp_p\bigg(\sum_{j=1,j\neq
m}^{q-1}h_{x}^{(j)}\bigg)+\theta}=
$$
$$\exp_p\bigg(\sum_{j=1,j\neq i}^{q-1}h_{x}^{(j)}\bigg), \eqno(3.8)
$$

here we have used the notation
$h_x=(h_x^{(1)},\cdots,h_x^{(q-1)})$. Writing
$h_x^{(i)'}=\sum_{j=1,j\neq i}^{q-1}h_{x}^{(j)}$  we immediately
get (3.3) from (3.8).

{\it Sufficiency.} Now assume that (3.3) is valid, then it implies
(3.8), and hence (3.7). From (3.7)  we obtain the following
equality $$ a(x)\exp_p\bigg(h_x\s_i\bigg)=\prod_{y\in
S(x)}\sum_{\s(y)\in\Psi}
\exp_p\bigg(J\delta_{\s_i,\s(y)}+h_y\s(y)\bigg), \ \ i=1,2,...,q.
$$ This equality implies $$ \prod_{x\in
W_{n-1}}a(x)\exp_p\bigg(h_x\s(x)\bigg)=$$ $$\prod_{x\in
W_{n-1}}\prod_{y\in S(x)}\sum_{\s(y)\in\Psi}
\exp_p\bigg(J\delta_{\s(x),\s(y)}+h_y\s(y)\bigg), \ \ i=1,2,...,q,
\eqno(3.9) $$ where $$ \s(z)=\left\{ \ba{ll} \s_i, \ \ z=x\\
\s(z), \ \ z\neq x\\ \ea \right. \ \ i=\overline{1,q}. $$

Writing $A_n(x)=\prod_{x\in W_n}a(x)$ we find from (3.9) and (3.2)
$$
Z_{n-1}A_{n-1}\m^{(n-1)}(\s_{n-1})=Z_n\sum_{\s^{(n)}}\m^{(n)}(\s_{n-1},\s^{(n)})
$$
Since each $\m^{(n)}$, $n\geq 1$  is a $p$-adic probability
measure,   we  have
$$
\sum_{\s_{n-1}}\sum_{\s^{(n)}}\m^{(n)}(\s_{n-1},\s^{(n)})=1, \ \
\sum_{\s_{n-1}}\m^{(n-1)}(\s_{n-1})=1.
$$
Therefore from these equalities we find $Z_{n-1}A_{n-1}=Z_n$ which
means that (3.2) holds.

Observe that according to this Theorem  the problem of describing
of $p$-adic Gibbs measures reduces to the describing of solutions
of the equation (3.3).

\section{The problem of phase transitions}

\subsection{The existence of phase transition for the $p$-adic Potts model}

Write
$$
\Lambda=\{ h=(h_x\in \Q_p^{q-1}, x\in V) : h_x \ \
\textrm{satisfies the equation (3.3)}\}.
$$

To prove the existence of phase transition it suffices to show
that there are two different sets of vectors in $\Lambda$. The
description of  arbitrary elements of the set $\Lambda$ is a
complicated problem.

In this paper we restrict ourselves to the description of
translation - invariant  elements of $\Lambda$, in which $h_x=h$
is independent of $x$.

Let $h_x=h=(h_1,...,h_{q-1})$ for all $x\in V$. Then (3.3) implies
$$
\exp_p(h_i)=\biggl(\frac{(\theta-1)\exp_p(h_i)+\sum_{j=1}^{q-1}\exp_p(h_j)+1}
{\sum_{j=1}^{q-1}\exp_p(h_j)+\theta}\biggr)^k, \ \ i=1,2,...,q-1.
\eqno(4.1) $$ Observe that for every $i=1,2,...,q-1$  $h_i=0$
satisfies $i$-th equation. Substituting $h_j=0$ at $j=2,3,...,q-1$
and writing $z=\exp_p(h_1)$ from the first  equation of (4.1)  $$
z=\biggl(\frac{\theta z+q-1}{z+\theta+q-2}\biggr)^k. \eqno(4.2) $$
Put $$ A=\theta z+q-1, \ \ \ B=z+\theta+q-2. $$ Then by (4.2) $$
B^k(z-1)=(\theta-1)(z-1)(A^{k-1}+A^{k-2}B+...+B^{k-1}). $$ Hence
if $z\neq 1$  $$ B^k=(\theta-1)(A^{k-1}+A^{k-2}B+...+B^{k-1}).
\eqno(4.3) $$ Using Lemma 2.2 it is easy to see that $$ |A|_p
\left\{ \ba{ll} \leq \dsf{1}{p}, \ \ \textrm{if \  $q\in
p\mathbb{N}$},\\[2mm] = 1, \ \ \textrm{if \ $q\notin
p\mathbb{N}$},\\ \ea \right. \ \ p\geq 3, $$ $$ |B|_p \left\{
\ba{ll} \leq \dsf{1}{p}, \ \ \textrm{if \  $q\in
p\mathbb{N}$},\\[2mm] = 1, \ \ \textrm{if \ $q\notin
p\mathbb{N}$},\\ \ea \right. \ \ p\geq 3, $$ $$ |A|_2 \left\{
\ba{lll} \leq \dsf{1}{4}, \ \ \textrm{if \  $q\in
2^2\mathbb{N}$},\\[2mm] =\dsf{1}{2}, \ \ \textrm{if \ $q\in
2\mathbb{N}\setminus 2^2\mathbb{N}$},\\[2mm] = 1, \ \ \textrm{if \
$q\notin 2\mathbb{N}$},\\ \ea \right. \ \ p=2. $$ $$ |B|_2 \left\{
\ba{lll} \leq \dsf{1}{4}, \ \ \textrm{if \  $q\in
2^2\mathbb{N}$},\\[2mm] = \dsf{1}{2}, \ \ \textrm{if \ $q\in
2\mathbb{N}\setminus 2^2\mathbb{N}$},\\[2mm] = 1, \ \ \textrm{if \
$q\notin 2\mathbb{N}$},\\ \ea \right. \ \ p=2. $$

From these inequalities we get

1) If $p=2$; $q\in 2\mathbb{N}\setminus 2^2\mathbb{N}$ we have
\bea |B|_2=|A|_2=\dsf{1}{2}, \ \
|B|_2^k=\dsf{1}{2^k}>\dsf{1}{4}\cdot\dsf{1}{2^{k-1}} \nonumber\\
|\theta-1|_2|A^{k-1}+...+B^{k-1}|_2\leq
\dsf{1}{4}\cdot\dsf{1}{2^{k-1}}. \nonumber \eea Here we have used
the strong triangle inequality and $|\theta-1|_2\leq \dsf{1}{4}$
(see Lemma 2.2). From the  last inequalities we infer that the
equation (4.3) has no solution.

2) If $p\geq 3$, $q\notin p\mathbb{N}$ (resp. $q\notin
2\mathbb{N}$ if $p=2$) then $$ |B|_p=|A|_p=1, \ \
|\theta-1|_p|A^{k-1}+...+B^{k-1}|_p\leq \dsf{1}{p}. $$ Hence in
this case the equation (4.3) has no solution either.

3) If $p\geq 3$, $q\in p\mathbb{N}$ (resp. $p=2$, $q\in
2^2\mathbb{N}$) then it is easy to see that the equation (4.3) may
have a solution.

Thus we have proved the following

{\bf Theorem 4.1.} {\it If $p\geq 3$, $q\in p\mathbb{N}$ (resp.
$p=2$, $q\in 2^2\mathbb{N}$) then the equation (4.1) may has at
least two solutions for every $k\geq 1$.}

According to Theorem 4.1 in the sequel we will assume that $p\geq
3$, $q\in p\mathbb{N}$ (resp. $p=2$, $q\in 2^2\mathbb{N}$). Note,
that in case of $k>2$ the problem of description of the solutions
of (4.3) becomes difficult. For simplicity we restrict ourselves
to the case $k=2$. Then (4.3) has the form $$
z^2+(2\theta-\theta^2+2q-3)z+(q-1)^2=0. \eqno(4.4) $$ Observe that
the solution of (4.4) can be written by $$
z_{1,2}=\frac{-(2\theta-\theta^2+2q-3)\pm(\theta-1)\sqrt{\theta^2-2\theta+5-4q}}{2}.
\eqno(4.5) $$ We must check the existence of
$\sqrt{\theta^2-2\theta+5-4q}$ and additionally the inequality
$|z_{1,2}-1|_p<p^{-1/(p-1)}$ which is equivalent to the condition
$|h_x|_p<p^{-1/(p-1)}$.

From $|2q-(\theta-1)^2|_p<p^{-1/(p-1)}$ we find
$|\theta-1|_p<p^{-1/(p-1)}$ for every prime number $p$. It then
follows from (4.5) that $|z_{1,2}-1|_p<p^{-1/(p-1)}$.

Now we check the existence of  $\sqrt{(\theta-1)^2+4(1-q)}$ . We
use canonical form of $p$-adic  numbers (2.1).

1) Let $p=2$. Then $|\theta-1|_2=|J|_2\leq\dsf{1}{4}$, hence
$\theta-1=2^{\g}\varepsilon$, $\g\geq 2$, $|\varepsilon|_2=1. $
According to our assumption $q=2^{2+m}s$, $m\geq 0$, $(s,2)=1$,
whence we can write $s=\sum_{i=0}^lc_i2^i$, $c_0=1$,
$c_i\in\{0,1\},i=1,2,...,l$. Hence we have $$
(\theta-1)^2+4(1-q)=2^2-2^{4+m}s+2^{2\g}\varepsilon^2=$$ $$
=2^2(1+\sum_{i=0}^l(2-c_i)2^{2+m+i}+2^{2\g-2}\varepsilon^2).
\eqno(4.6) $$

2) Let $p=3$. Then $q=3^ms$, $(s,3)=1$,$m\geq 1$, whence
$4s=\sum_{i=0}^lb_i3^i$, $b_0=1,2$,
$b_i\in\{0,1,2\},i=\overline{1,l}$. The inequality
$|\theta-1|_3\leq\dsf{1}{3}$ implies that
$\theta-1=3^{\g}\varepsilon$, $\g\geq 1$, $|\varepsilon|_3=1.$
Hence we get $$
(\theta-1)^2+4(1-q)=1+3+\sum_{i=0}^l(3-b_i)3^{m+i}+3^{2\g}\varepsilon^2.
$$

3) Let $p\geq 5$. Then $q=p^ms$, $(s,p)=1$,$m\geq 1$, whence
$4s=\sum_{i=0}^lb_ip^i$, $b_0=1,2,...,p-1$, $b_i=0,1,...,p-1$,
$i=1,2,...,l$. The inequality $|\theta-1|_p\leq\dsf{1}{p}$ implies
that $\theta-1=p^{\g}\varepsilon$, $\g\geq 1$,
$|\varepsilon|_p=1.$ Hence we get $$
(\theta-1)^2+4(1-q)=4+\sum_{i=0}^l(p-b_i)p^{m+i}+p^{2\g}\varepsilon^2.
$$

 We can now check all conditions of  Lemma 2.1:
observe that the each case the first condition of Lemma 2.1 is
fulfilled. It remains to check the second condition, i.e. the
equation $x^2\equiv a_0(\textrm{mod $p$})$ has solution $x\in\Z$.

1) Let $p=2$. In this case $a_0=1$, then it is easy to see that
the equation $x^2\equiv 1(\textrm{mod $2$})$ has solution
$x=2N+1$, $N\in \Z$. Besides it must be $a_1=a_2=0$. From (4.6)
one can find that
$$
a_1=a_2=0 \ \ \textrm{if and only if either} \ m=0,\g=2 \ \
\textrm{or} \ m>1,\g>2.
$$
Thus in this case for $2$-adic Potts model a phase transition
occurs.

2) Let $p=3$. In this case $a_0=1$, then it is not difficult to
check that the equation $x^2\equiv 1(\textrm{mod $3$})$ has the
solution $x=3N+1$, $N\in \Z$.

3) Let $p\geq 5$. In this case $a_0=4$, then $x^2\equiv
2(\textrm{mod $p$})$ has the solution $x=pN+2$, $N\in \Z$.

Consequently we have proved the following

{\bf Theorem 4.2.} {\it i) Let $p=2$,$q\in 2^2\mathbb{N}$ and
$J\neq 0$. If $q=2^2s$, $(s,2)=1$, $|J|_2=\dsf{1}{4}$ or
$q=2^ms$,$m\geq 3$,$(s,2)=1$, $|J|_2\leq\dsf{1}{4}$ then there
exists a phase transition for the $2$-adic Potts model (2.4) on a
Cayley tree of order 2.

ii) Let $p\geq 3$, $q\in p\mathbb{N}$, and $0<|J|_p\leq\dsf{1}{p}$
then there exists a phase  transition for $p$-adic Potts model on
a Cayley tree of order 2.}\\

Observe that if $q=2$ then the Potts model becomes the Ising
model, so from this theorem and Theorem 4.1 we have the following

{\bf Corollary 4.3.} {\it Let $k\geq 1$. Then for the $p$-adic
Ising  model on the Cayley tree of order $k$ there is no phase
transition.}

{\bf Conjecture.} {\it Let all conditions of Theorem 4.2 be
satisfied. Then there is a phase transition for the $p$-adic Potts
model on the Cayley tree of order $k\ (k\geq 3)$.}

Now we investigate when the $p$-adic Gibbs measure with the
solutions (4.2) is bounded.

{\bf Theorem 4.4.}  {\it The $p$-adic Gibbs measure $\m$ for the
$p$-adic Potts model on the Cayley tree of order $k$ is bounded if
$q\notin p\mathbb{N},$ otherwise it is not bounded.}

{\bf Proof.} To prove the assertion of theorem it suffices to show
that the values of $\m$ on cylindrical subsets are bounded. We
estimate $|\m^{(n)}(\s_n)|_p$: \bea
|\m^{(n)}(\s_n)|_p=\bigg|\frac{\exp_p\{\tilde H(\s_n)\}}
{\sum_{\tilde\s_n\in \O_{V_n}}\exp_p\{\tilde
H(\s_n)\}}\bigg|_p=\nonumber\\ \frac{1}{\bigg|\sum_{\tilde\s_n\in
\O_{V_n}}(\exp_p\{\tilde H(\s_n)\}-1)+q^{V_n}\bigg|_p}=1\nonumber
\eea if $q\notin p\mathbb{N}$. Here $$ \tilde
H(\s_n)=H(\s_n)+\sum_{x\in W_n}h_*\s(x), $$ and $h_*$ is a
solution of (4.4), and we have used $|\exp_p\{\tilde
H(\s_n)\}-1|_p\leq\dsf{1}{p}$. Note that $h_*$ is a vector which
has the form $h_*=(\underbrace{h,0,...,0}_{q-1})$.

Write $$
p_{ij}=\frac{\exp_p(J\delta_{ij}+h_*(i+j))}{\sum_{km}\exp_p(J\delta_{km}+h_*(k+m))},
$$ where $i,j\in\{\s_1,...,\s_{q-1}\}$.

To prove that the measure $\m$ is not bounded at $q\in
p\mathbb{N}$ it is enough to show that its marginal (formdary)
measure is not bounded.  Let $\pi=\{...,x_{-1},x_0,x_1,...\}$ be
an arbitrary infinite path in $\Gamma^k$. From (3.1) one can see
that a marginal (formdary) measure $\m_{\pi}$ on $\Psi^{\pi}$ has
the form $$
\m_{\pi}(\omega_n)=p_{\omega(x_{-n})}\prod_{m=-n}^{n-1}p_{\omega(x_m)\omega(x_{m+1})},
\eqno(4.7) $$ here $\omega_n :\{x_{-n},...,x_0,...,x_n\}\to \Psi$,
i.e. $\omega_n$ is a configuration on
$\{x_{-n},...,x_0,...,x_n\}$, $p_i$ is an invariant vector for the
matrix $(p_{ij})_{ij=1}^{q}$.

Using (2.3) and the form of $h_*$ we have $$
|p_{ij}|_p=\frac{1}{\bigg|2\exp_p(J+2h)+(q-2)\exp_p(J)+
q\sum_{i,j,\i\neq j}\exp_p(h_*(i+j))\bigg|_p}= $$ $$
=\frac{1}{\bigg|2(\exp_p(J+2h)-1)+2+(q-2)\exp_p(J)+q\sum_{i,j,i\neq
j}\exp_p(h_*(i+j))\bigg|_p}\geq p. \eqno(4.8) $$ for all $i,j$.
Here we have used Lemma 2.2 and $|q|_p\leq\dsf{1}{p}$.  From (4.7)
and (4.8) we find that $\m_{\pi}$ is not bounded. Hence the
theorem is proved.\hfill $\square$

{\bf Corollary 4.5.} {\it The $p$-adic Gibbs measure $\m$
corresponding to the $p$-adic Ising model on the Cayley tree of
order $k$ is bounded if $p\neq 2$, otherwise it is not bounded.}

{\bf Remark.} From Theorems 4.2 and 4.4 we see that a phase
transition occurs when  $p$-adic Gibbs measures are not bounded.
For the $p$-adic Ising model we  know that a phase transition does
not occur, so corollary 4.5 implies that if $p=2$ even in this
case the $p$-adic Gibbs measure may not be bounded.

\subsection{The uniqueness of Gibbs measure for the $p$-adic Potts model}

If $q\in p\mathbb{N}$ then the equation (4.3) may have  two
solutions. But thus remains the case $q\notin p\mathbb{N}$ and a
question naturally arises: is there a phase transition in this
case or not? In this section we will prove the uniqueness of
$p$-adic Gibbs measure for the $p$-adic Potts model in that case.

Let us first prove some technical results.

{\bf Lemma 4.6.} {\it If $|a_i-1|_p\leq M$ and $|a_i|_p=1$,
$i=1,2,...,n$, then $$ \bigg|\prod_{i=1}^{n}a_i-1\bigg|_p\leq M.
\eqno(4.9) $$ }

{\bf Proof.} We prove this by induction on $n$. The case $n=1$ is
nothing but the condition of lemma. Suppose that (4.9) is valid
for $n=m$. Now let $n=m+1$. Then we have \bea
\bigg|\prod_{i=1}^{m+1}a_i-1\bigg|_p=\bigg|\prod_{i=1}^{m+1}a_i-\prod_{i=1}^ma_i+\prod_{i=1}^ma_i-1\bigg|_p\leq\nonumber\\
\leq\max\bigg\{\bigg|\prod_{i=1}^ma_i(a_{m+1}-1)\bigg|_p,
\bigg|\prod_{i=1}^ma_i-1\bigg|_p\bigg\}\leq M\nonumber \eea

This completes the proof.\hfill $\square$

{\bf Lemma 4.7.} {\it Let $u_i=\prod\limits_{j=1,j\neq
i}\exp_p(h_j)$, where $h=(h_1,...,h_{q-1})$, $\|h\|_p\leq
\dsf{1}{p}$, then $|u_i|_p=1$ and $$ |u_i-1|_p\leq\dsf{1}{p}, $$
for all $i=1,...,q-1$.}

{\bf Proof.} From Lemma 2.2 we infer that $|u_i|_p=1$, since
$|\exp_p(h_i)-1|_p\leq \dsf{1}{p}$ and $|\exp_p(h_i)|_p=1$. So all
conditions of Lemma 4.6 are satisfied, hence we get
$|u_i-1|_p\leq\dsf{1}{p}.$ \hfill $\square$\\[1mm]

Write (see Theorem 3.3) $$ U_i(h,\theta,q)=\exp_p(F(h,\theta,q))=
\prod_{j=1,j\neq
i}^{q-1}\frac{(\theta-1)u_j+\sum_{j=1}^{q-1}u_j+1}
{\sum_{j=1}^{q-1}u_j+\theta}, \eqno(4.10) $$ where $i=1,...,q-1$.
For brevity  we use $U_i$ instead of $U_i(h,\theta,q)$.

{\bf Lemma 4.8.} {\it Let $q\notin p\mathbb{N}$, then $|U_i|_p=1$
and
$$
|U_i-1|_p\leq \dsf{1}{p}\|h\|_p  \ \ \textrm{for} \ \
i=1,2,...,q-1 \eqno(4.11)
$$
}

{\bf Proof.} Put $$ K_i=\frac{(\theta-1)u_i+\sum_{j=1}^{q-1}u_j+1}
{\sum_{j=1}^{q-1}u_j+\theta}. $$ We compute the norm of $K_i$.
From $|\theta-1|_p\leq\dsf{1}{p}$, $|q|_p=1$ and Lemma 4.7 we
obtain $$
|K_i|_p=\bigg|\frac{(\theta-1)u_i+\sum_{j=1}^{q-1}(u_j-1)+q}
{\sum_{j=1}^{q-1}(u_j-1)+(\theta-1)+q}\bigg|_p=1. $$ Here we have
used the strong triangle property of the norm $|\cdot |_p$.
Observe that $U_i=\prod\limits_{j=1,j\neq i}^{q-1}K_j$, and hence
$|U_i|_p=1$.

Now estimate $|K_i-1|_p$: $$
|K_i-1|_p=\bigg|\frac{(\theta-1)(u_i-1)}
{\sum_{j=1}^{q-1}(u_j-1)+(\theta-1)+q}\bigg|_p\leq
\dsf{1}{p}\|h\|_p. $$ Consequently, we find that the conditions of
Lemma 4.6 are satisfied for $K_i$, $i=1,2,...,q-1$, whence
(4.11).The lemma
 is proved.

Wtite $$ R_i(h_x,\theta,q)=\prod_{y\in S(x)}U_i(h_y,\theta,q), \ \
x\in V. \eqno(4.12) $$ Using Lemmas 4.6 and 4.8 one proves the
following

{\bf Lemma 4.9.} {\it Let $q\notin p\mathbb{N}$, then $$
|R_i(h_x,\theta,q)-1|_p\leq\dsf{1}{p}\max_{y\in S(x)}\|h_y\|_p. $$
}

{\bf Corollary 4.10.} {\it Let  $q\notin p\mathbb{N}$, then
$$
\|h_x\|_p\leq\dsf{1}{p}\max_{y\in S(x)}\|h_y\|_p, \ \ x\in V.
\eqno(4.13)
$$
}

{\bf Proof.} From (3.7) and (4.12) one can see that
$\exp_p(h^x_i)=R_i(h_x,\theta,q)$. Hence using Lemma 2.2 and Lemma
4.9 we infer that $$
|h^x_i|_p=|R_i(h_x,\theta,q)-1|_p\leq\dsf{1}{p}\max_{y\in
S(x)}\|h_y\|_p, $$ whence (4.13).\hfill $\square$ \\[1mm]

Now we can   formulate the main result of this subsection.

{\bf Theorem 4.11.} {\it Let $k\geq 1$ and $q\notin p\mathbb{N}$,
$|J|_p\leq \dsf{1}{p}$ and $p$ be any prime number. Then for the
$p$-adic Potts model (2.4) on the Cayley tree of order $k$ there
is no phase transition.}

{\bf Proof.} To prove it enough to show that $\Lambda =\{h_x\equiv
0\}$. In order to do this we will show that for arbitrary
$\varepsilon>0$ and every $x\in V$ we have $\|h_x\|_p<\varepsilon$
. Let $n_0\in\mathbb{N}$ be such that
$\dsf{1}{p^{n_0}}<\varepsilon$. According to Corollary 4.10 we
have \bea \|h_x\|_p\leq\dsf{1}{p}\|h_{x_{i_0}}\|_p\leq \nonumber\\
\leq \dsf{1}{p^2}\|h_{x_{i_0,i_1}}\|_p\leq\cdots
\leq\dsf{1}{p^{n_0-1}}\|h_{x_{i_0,...,i_{n_0-2}}}\|_p\leq\dsf{1}{p^{n_0}}<\varepsilon,\nonumber
\eea here $x_{i_0,...,i_n,j}$, $j=1,2,...,k$ are direct successors
of $x_{i_0,...,i_n}$, where
$\|h_{x_{i_0,...,i_m}}\|_p=\max\limits_{1\leq j\leq
k}\{\|h_{x_{i_0,...,i_{m-1},j}}\|_p\}$. This completes the proof.
\hfill $\square$

 {\bf Remark.} If $q\notin p\mathbb{N}$ then
theorem 4.4 says that the $p$-adic Gibbs measure corresponding to
the Potts model is bounded, hence by Theorem 4.11 we see that in
this case  there is only bounded $p$-adic Gibbs measure for the
$p$-adic Potts model.
\\

{\bf Acknowledgement}
 This work was done within
the scheme of Mathematical fellowship at the Abdus Salam
International Center for Theoretical Physics (ICTP) and the
authors thank ICTP and IMU/CDE - program for providing financial
support and all facilities. The authors acknowledge with gratitude
to Professors I.V.Volovich and A.Yu.Khrennikov for the helpful
comments and discussions.

The authors is also grateful to the referee for useful
suggestions.

{\bf References}

1. I.V.Volovich, Number theory as the ultimate physical theory,
Preprint, TH, 4781/87.

2. I.V.Volovich, {\it $p$-adic strings, Class. Quantum Grav.} {\bf
4}(1987) L83-L87.

3. P.G.O.Freund, and E.Witten, {\it Adelic string amplitudes,
Phys. Lett.} {\bf B199}(1987) 191-194.

4. E.Marinary and G.Parisi, {\it On the $p$-adic five point
function, Phys.Lett.} {\bf 203B}(1988) 52-56.

5. I.Ya.Araf'eva, B.Dragovich, P.H.Frampton and I.V.Volovich, {\it
Wave function of the universe and $p$-adic gravity, Int. J.
Mod.Phys. A} {\bf 6}(1991) 4341-4358.

6. A.Yu.Khrennikov, {\it $p$-adic quantum mechanics with $p$-adic
valued functions, J.Math.Phys.} {\bf 32}(1991) 932-936.

7. V.S.Vladimirov and I.V.Volovich and E.I.Zelenov, {\bf  $p$-adic
Analysis and Mathematical Physics} (World Scientific, 1994).

8. A.Yu.Khrennikov, {\bf $p$-adic Valued Distributions in
Mathematical Physics} (Kluwer, 1994).

9. A.N.Kolmogorov, {\bf  Foundation of the Theory of Probability}
(Chelsea, 1956).

10. A.Yu.Krennikov, {\it  $p$-adic valued probability measures,
Indag. Mathem. N.S.} {\bf 7}(1996) 311-330.

11. A.Yu.Krennikov, S.Ludkovsky {\it  On infinite products of
non-Archimedean measure spaces, Indag. Mathem. N.S.} {\bf
13}(2002) 177-183.

12. A.Yu.Khrennikov and S.Ludkovsky {\it Stochastic process on
non-Archimedean spaces with values in non-Archimedean fields, Adv.
Stud. in Contemp. Math.} {\bf 5}(2002), 1, 57-91.

13. N.N.Ganikhodjaev, F.M.Mukhamedov and U.A.Rozikov, {\it
Existence of a phase transition for the Potts $p$-adic model on
the set $\mathbb{Z}$, Theor.Math.Phys.} {\bf 130}(2002) 425-431.

14 D.Ruelle, {\bf  Statistical Mechanics: Rigorous Results}
(Benjamin, 1969).

15. N.Koblitz, {\bf  $p$-adic numbers, $p$-adic analysis and
zeta-function} (Springer, 1977).

16. F.Q.Gouvea, {\bf $p$-adic numbers} (Springer, 1991).

\end{document}